# Pel,

## A Programming Language for Orchestrating AI Agents

Behnam Mohammadi*

June 08, 2025

**Abstract**

The proliferation of Large Language Models (LLMs) has opened new frontiers in computing, yet controlling and orchestrating their capabilities beyond simple text generation remains a challenge. Current methods, such as function/tool calling and direct code generation, suffer from limitations in expressiveness, scalability, cost, security, and the ability to enforce fine-grained control. This paper introduces Pel, a novel programming language specifically designed to bridge this gap. Inspired by the strengths of Lisp, Elixir, Gleam, and Haskell, Pel provides a syntactically simple, homoiconic, and semantically rich platform for LLMs to express complex actions, control flow, and inter-agent communication safely and efficiently. Pel's design emphasizes a minimal, easily modifiable grammar suitable for constrained LLM generation, eliminating the need for complex sandboxing by enabling capability control at the syntax level. Key features include a powerful piping mechanism for linear composition, first-class closures enabling easy partial application and functional patterns, built-in support for natural language conditions evaluated by LLMs, and an advanced Read-Eval-Print-Loop (REPeL) with Common Lisp-style restarts and LLM-powered helper agents for automated error correction. Furthermore, Pel incorporates automatic parallelization of independent operations via static dependency analysis, crucial for performant agentic systems. We argue that Pel offers a more robust, secure, and expressive paradigm for LLM orchestration, paving the way for more sophisticated and reliable AI agentic frameworks.

*Ph.D., Carnegie Mellon University, behnamm@alumni.cmu.edu

# Contents

# 1 Introduction

The capabilities of Large Language Models (LLMs) have expanded dramatically, moving beyond simple text generation towards executing complex tasks and interacting with external systems. A critical challenge in this evolution is enabling LLMs to perform actions reliably, safely, and expressively. Current industry approaches primarily fall into two categories, each with significant drawbacks.

The first, widely adopted approach is **function calling** or **tool calling** ("Function calling and other API updates," 2024). Here, programmers pre-define functions, expose their signatures (often as JSON Schema) to the LLM, and the model generates JSON payloads specifying which function to call with what arguments. While useful for simple tasks, this method suffers from several limitations. First, it struggles to represent complex control flow (conditionals, loops), sequential dependencies beyond simple chaining, or parallel execution patterns. In this method conditional logic—even a condition easily verifiable by code—often relies solely on the LLM's judgment (e.g., "call action A if condition X is met"), which not only can lead to potential inaccuracies, but also results in verification opacity. Moreover, this method does not scale in real-world scenarios: Defining hundreds of functions for complex agents becomes unmanageable, and the resulting large JSON schemas passed to the LLM increase costs and can significantly degrade the model's reasoning performance. Finally, function calling is a rigid method: Actions are limited strictly to the pre-defined functions. The LLM cannot compose existing tools in novel ways or perform computations not explicitly provided.

The second approach involves letting the LLM **generate code directly** in a general-purpose language like Python (e.g., ChatGPT Code Interpreter, "ChatGPT plugins," 2024). This offers greater flexibility and allows the LLM to leverage existing libraries. However, it introduces severe challenges. First, running arbitrary code generated by an LLM is inherently dangerous. Prompt injection attacks may trick the LLM into generating malicious code that can compromise the system, exfiltrate data, or perform unintended actions. Sandboxing helps but is complex and not foolproof. Secondly, restricting the LLM's capabilities within a powerful language like Python is extremely difficult. For example, preventing the use of loops or specific library functions requires sophisticated static analysis or runtime monitoring. Modifying the grammar of languages like Python to disable features at the generation level is impractical due to their complexity (e.g., Python's EBNF grammar spans hundreds of lines). Converting such large grammars to formats suitable for constrained generation (like regex) results in massive performance degradation.

This research stems from my decade-long fascination with programming language design, sparked by exploring both mainstream and esoteric languages and their unique ideas. During my PhD work on a project leveraging LLMs, the inadequacy of existing action-performing mechanisms became starkly apparent. We needed a way for an LLM to specify complex, conditional, and potentially concurrent actions safely and reliably, particularly in the context of orchestrating multiple AI agents. This led to the development of Pel, an intermediate lan-guage specifically designed to be written by LLMs and safely interpreted by the host system.

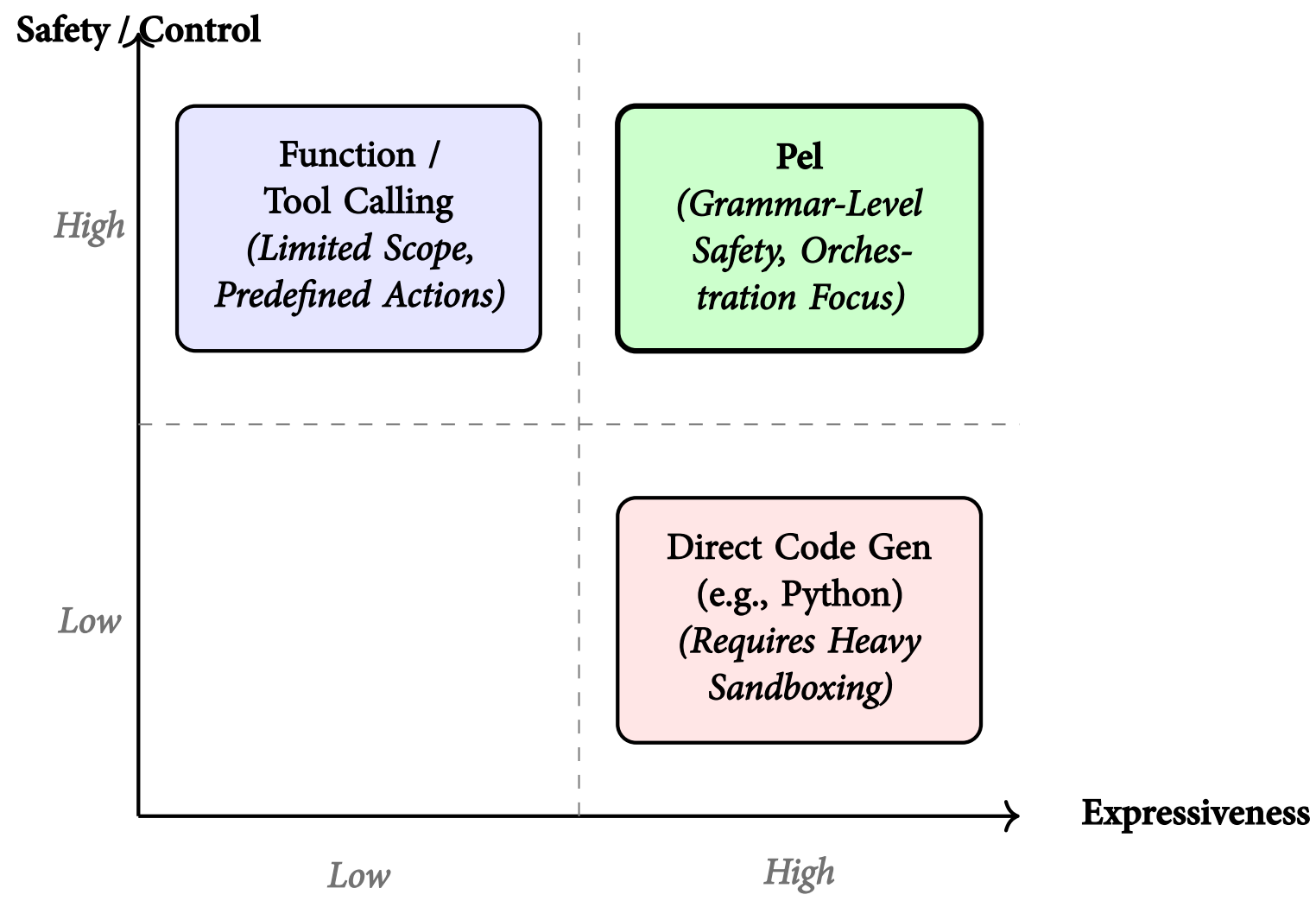

Figure 1: Pel's Position in the LLM Orchestration Landscape, balancing expressiveness and safety.

Pel combines a Lisp-like syntax with functional programming concepts, resulting in a language with several distinctive characteristics. First, it is safe by design—its simple, regular grammar can be easily modified to restrict LLM generation at the source level, disabling unwanted features (e.g., network access, file I/O, specific functions) with guarantees, eliminating the need for complex sandboxing. Second, Pel is highly expressive, supporting conditionals, loops (as non-strict functions), variable bindings, complex data structures, and powerful composition via pipes (See Figure 1). The language is also inherently LLM-friendly. Its uniform syntax and simple grammar facilitate easier learning (in-context) and reliable generation by LLMs. Its linear composition style (pipes) aligns well with the sequential token generation process of LLMs. Furthermore, Pel provides fine-grained control over execution, including automatic asynchronous execution for performance-critical agentic systems. Finally, it integrates natively with LLM capabilities for tasks like evaluating natural language conditions and providing automated debugging help within its interactive environment.

This paper details the design philosophy, syntax, semantics, and novel features of Pel. We elaborate on its suitability for LLM code generation and its interactive Read-Eval-Print-Loop (aptly named “REPeL”). We also briefly illustrate its application in orchestrating hierarchical AI agent systems, demonstrating Pel’s potential as a foundational technology for building sophisticated and reliable agentic AI systems.

## 2 Literature Review

The development of agent-based AI systems has seen significant advancement in recent years, particularly with the emergence of large language models (LLMs). Within this domain, researchers have explored various frameworks for agent architecture and orchestration, with systems like AutoGPT and BabyAGI representing early attempts at autonomous agentic systems (Weng, 2023). More formalized approaches have emerged, such as the ReAct framework, which combines reasoning and acting in language agents through a process of thought-action-observation cycles (Yao et al., 2022).

The challenge of reliably coordinating multiple agents has led to the development of multi-agent frameworks like AgentVerse (Chen et al., 2023), which provides a customizable platform for constructing and orchestrating agent societies in various application domains. Similarly, MetaGPT (Hong et al., 2023) proposes a meta-programming framework that enables collaborative problem-solving among multiple agents with specialized roles. These multi-agent systems often struggle with reliable inter-agent communication and coordination, with recent work by Park et al. highlighting challenges in agent-to-agent interaction patterns (Park et al., 2023).

Despite these advances, a critical gap exists in the orchestration of AI agents. Current approaches predominantly fall into two categories, each with significant limitations. The first approach is function calling or tool calling, which allows LLMs to interact with predefined functions but struggles with complex control flow, sequential dependencies, and scalability in real-world scenarios (IBM, 2025). As noted by Microsoft’s Magentic-One research, function calling becomes unwieldy when orchestrating numerous agents with complex interdependencies (Microsoft Research, 2024). The Berkeley Function-Calling Leaderboard highlights ongoing challenges in function calling reliability across different programming languages and API scenarios (Berkeley, n.d.).

Function calling approaches face several critical limitations: they struggle to represent complex control flow patterns (conditionals, loops), provide limited scaling capacity when defining large numbers of functions, and lack expressiveness for novel tool compositions (2024). Additionally, security concerns arise when verification is delegated solely to LLM judgment rather than programmatically verifiable code (BentoML, 2024). These limitations impede the development of sophisticated agentic systems capable of handling real-world tasks with appropriate governance mechanisms (OpenAI, 2024).

The second approach—allowing LLMs to generate code directly in general-purpose languages—offers greater flexibility but introduces severe security risks and challenges in capability restriction. Pel positions itself in this gap, providing a specialized language with a restricted grammar that enables safe yet expressive agent orchestration. Unlike function calling frameworks that limit expressivity, or general-purpose language generation that compromises safety, Pel offers a middle ground that addresses both concerns simultaneously.

Pel's approach to grammar-level safety leverages recent advances in constrained generation techniques. Research on grammar-constrained decoding demonstrates that formal grammars can successfully restrict LLM output to follow specific structures (Geng et al., 2023), ensuring syntactic validity without requiring fine-tuning. The approach of converting context-free grammars to regular expressions for constrained LLM generation has gained traction as demonstrated by tools like ReLLM (Rickard, 2024) and frameworks such as Domino (Wagner et al., 2024), which implement efficient and minimally-invasive constrained decoding.

These constrained generation approaches offer significant advantages over traditional sandboxing methods. As shown by Cooper (Cooper, 2024), constrained decoding guarantees valid outputs on first generation by restricting token distributions via state machines with regex or context-free grammars. This approach is particularly valuable for programming languages, where syntax errors can render generated code unusable. By enforcing grammar constraints at the generation level, Pel can guarantee the syntactic safety of LLM-generated code without compromising expressiveness.

On the programming language design front, a number of DSLs (Domain-Specific Languages) have been designed for specific AI tasks. DSPy (Khattab et al., 2023), for example, intro-duces a framework for programming foundation models that separates the optimization of prompts from their specification. Beyond such DSLs, though, general-purpose languages such as Lisp have had profound influence on AI, dating back to McCarthy's original work (McCarthy, 1960). Modern iterations of Lisp like Clojure have demonstrated the continued relevance of homoiconicity and S-expressions for representing and manipulating code as data (Hickey, 2008). Homoiconicity—"the ability to treat code as data" ("Exploring the Power of Artificial Intelligence in Lisp Programming," 2024)—facilitates metaprogramming and makes it particularly suitable for AI applications where programs need to be generated or manipulated by other programs (SIGPLAN Blog, 2020).

Pel's error handling system draws inspiration from Common Lisp's condition system, which has been acknowledged as a sophisticated mechanism for error recovery and program resilience (Pitman, 1988). Elixir and Gleam, which inform Pel's pipe operator syntax, demonstrate how functional language constructs can enhance code readability and composition (Valim, 2013). The pipe operator ( ▷ ) enables sequential data transformation that aligns well with LLMs' token-by-token generation pattern. Additionally, Common Lisp's error handling system, which influences Pel's REPeL design, provides sophisticated mechanisms for error recovery (Pitman, 1988) that are particularly valuable when dealing with potentially faulty LLM-generated code.

Pel's unique contribution lies in its synthesis of these various influences into a cohesive language specifically designed for LLM orchestration, addressing limitations in both function calling approaches and unrestricted code generation while providing its own novel, expressive, safe, and LLM-friendly programming environment for agentic systems.

# 3 Design Philosophy

Pel is designed with simplicity, regularity, and expressiveness in mind, tailored for generation by LLMs and safe interpretation. In particular, Pel strives for:

1. **Simplicity and Consistency:** Employ a minimal, regular syntax based on Lisp's S-expressions. This uniformity makes the language easier for LLMs to learn (better in-context learning) and parse.

2. **Grammar-Level Safety:** Define a concise grammar (expressible easily in EBNF) that can be readily modified. This allows developers to enable or disable language features (specific functions, control flow constructs, network access) *at the grammar level*. By using constrained generation techniques (e.g., regex sampling guided by the grammar), we can *guarantee* that the LLM cannot generate forbidden code paths, eliminating the need for runtime sandboxing for many security concerns (See Figure 2).
3. **Expressiveness for Orchestration:** Include essential control flow structures and data manipulation capabilities as first-class citizens within the language.
4. **Composable Linearity:** Provide mechanisms (like pipes) that allow LLMs to build complex workflows step-by-step without needing to plan the entire structure in advance or backtrack during generation.
5. **Seamless LLM Integration:** Natively incorporate mechanisms for leveraging LLM capabilities where appropriate (e.g., evaluating natural language conditions).
6. **Developer Ergonomics:** Offer an interactive development experience with robust error handling and debugging aids (the REPeL).

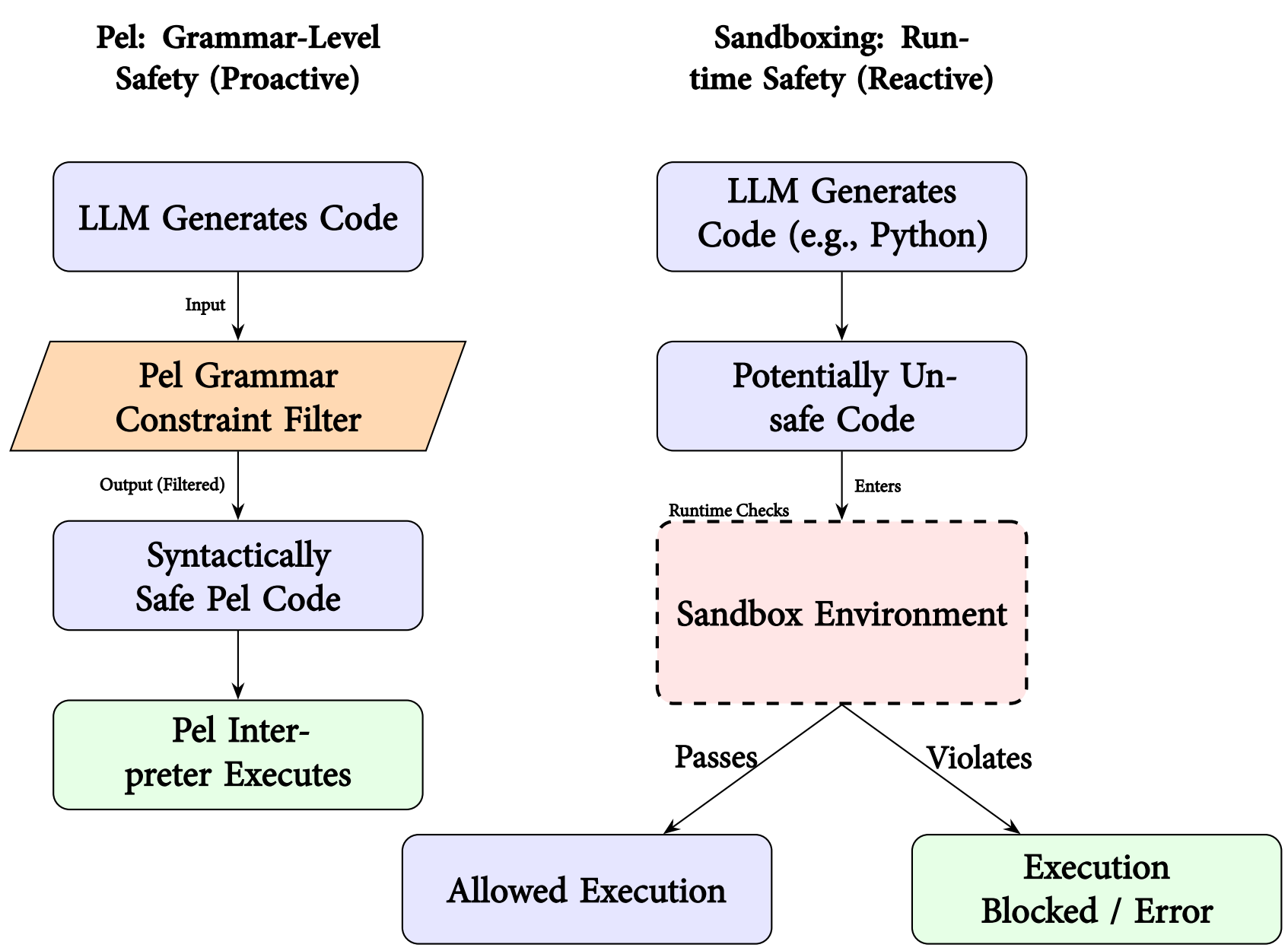

Figure 2: Comparison of Safety Mechanisms: Pel's Proactive Grammar Constraint vs. Reactive Runtime Sandboxing.

The design of Pel did not occur in a vacuum. It stands on the shoulders of giants in programming language history and draws inspiration from several key languages that have fascinated me over the years:

**Lisp:** The most apparent influence is Lisp, particularly its S-expression syntax `(...)`. This provides a simple, uniform structure (homoiconicity) that is easy to parse and manipulate programmatically. However, Pel diverges significantly from traditional Lisps. It avoids cons cells as the fundamental list structure in favor of distinct parenthesized evaluation lists `(...)` and bracketed literal lists `[...]`. Crucially, Pel enforces a strict operator-first interpretation for `(...)` forms and eliminates special forms entirely, opting for a unified function model.

**Elixir and Gleam:** The pipe operator (`▷`) is directly inspired by Elixir, providing a linear and readable way to compose functions. Pel extends this concept by using the `^` caret symbol for injecting the piped value into specific argument positions, offering greater flexibility than Elixir's default first-argument injection. This linearity is particularly advantageous for LLMs, which generate code sequentially.

**Haskell:** Pel incorporates automatic partial application, a hallmark of Haskell and other ML-family languages. When a function (or `PelClosure`) is called with fewer arguments than it expects, it automatically returns a new closure capturing the provided arguments, rather than raising an error. This is possible because Pel functions have a fixed arity (no variadic arguments), allowing the interpreter to "know" when a function is full and ready to fire. This simplifies the creation of higher-order functions and functional composition patterns. Pel's emphasis on immutability also echoes functional programming principles championed by Haskell.

**Common Lisp:** The design of Pel's interactive environment, the REPeL, draws inspiration from the powerful condition and restart system of Common Lisp. The ability to intercept errors, inspect the context, and choose how to proceed (e.g., retry, abort, provide a value, rewrite code) offers a much more robust development experience compared to typical REPLs, especially relevant when dealing with potentially faulty code generated by LLMs.

Pel aims to provide a unique and effective solution by synthesizing these influences within the specific context of LLM interaction while offering its own novel contributions.

# 4 The Pel Language

Pel is designed to be minimal yet expressive. Its core components are described below.

## 4.1 Syntax and Grammar

Pel uses a Lisp-like syntax based on S-expressions. Code consists of atoms and lists.

- **Atoms:** Basic indivisible values like numbers (`PelNum`), strings (`PelString`), booleans (`PelBool`), the null value (`PelNil`), symbols (`pelSymbol`), and keywords (`PelKey`).
- **Lists:** Sequences of elements enclosed in delimiters. Pel distinguishes between:
  - **Parenthesized Lists `()`:** Interpreted as expressions. The first element is *always* treated as the operator (a function/closure to be called), and the remaining elements are its arguments. Example: `(+ 1 2)`.
  - **Bracketed Lists `[]`:** Interpreted as literal data lists (`PelListLiteral`). Elements are evaluated, but the list itself is treated as data, not a function call. Example: `[1 #t "hello"]`. This avoids the ambiguity present in some Lisps where `()` can mean both function call and data list. Notice that literal lists are heterogeneous, unlike Python lists that can only contain one type of data.

This strict interpretation of `()` simplifies parsing and evaluation logic. The grammar is designed to be small and regular, making it amenable to constrained generation.

### 4.1.1 Pel EBNF Grammar

```
{(* Entry point: A program is zero or more expressions *)}
program = { expression } ;
(* An expression is a primary expression, potentially chained with pipes *)
expression = primary , { PIPE , primary } ;
(* A primary expression is the base unit before considering pipes *)
primary = atom
        | list
        | literal_list
        | quoted_expression
```

```
        ;
(* Atomic literal values *)
(* Note: KEY is included here. The parser interprets it contextually *)
(*       as either a literal key or the start of a key-value pair. *)
(*       The grammar allows KEY to be followed by another expression *)
(*       within lists or at the top level, covering the pair syntax. *)
atom = BOOL | NIL | NUMBER | STRING | SYMBOL | KEY ;
(* Standard parenthesized list *)
list = LPAREN , { expression } , RPAREN ;
(* Bracketed literal list *)
literal_list = LBRACKET , { expression } , RBRACKET ;
(* Quoted expression *)
(* The parser ensures the content is interpreted with allow_pairs=False *)
quoted_expression = QUOTE , expression ;
(* Terminal Symbols (provided by the tokenizer) *)
LPAREN   = "(" ;
RPAREN   = ")" ;
LBRACKET = "[" ;
RBRACKET = "]" ;
QUOTE    = "'" ;
PIPE     = "▷" ;
BOOL     = ? "#t" | "#f" ?
; (* Based on regex r"#t|#f" *)
NIL      = "#nil"
; (* Based on regex r"#nil" *)
STRING   = ? C-style string literal ?
; (* Based on regex r'"[^"]*"' *)
KEY      = ? keyword starting with ':' ?
; (* Based on regex r":[a-zA-Z0-9_\-+*/\\?!<>=.]+" *)
NUMBER   = ? integer or float literal ?
; (* Based on regex r"-?\d+(\.\d+)?" *)
SYMBOL   = ? sequence of non-whitespace, non-delimiter chars ?
; (* Based on regex r"[^\s()\[\]\";|>]+" *)
(* Ignored Tokens: comment, whitespace *)
```

The simplicity of this grammar is crucial. It allows for easy conversion to regular expressions for use in LLM constrained generation frameworks (like Guidance or LMQL). Developers can easily modify this EBNF (e.g., remove the `PIPE` rule, remove specific `SYMBOL`s corresponding to built-in functions) to create variants of Pel with restricted capabilities, enforced at the generation stage.

## 4.2 Data Types

Pel supports a range of fundamental data types:

- `PelNum`: Represents integer and floating-point numbers.
- `PelString`: Represents UTF-8 text strings, enclosed in double quotes (`"`).
- `PelBool`: Represents boolean values `#t` (true) and `#f` (false).
- `PelNil`: Represents the null or absence of value, denoted `#nil` or `()`. Importantly, unlike Python's `None`, `PelNil` does *not* evaluate to false in conditional contexts. Its truthiness is undefined.
- `PelListLiteral`: Represents 1-indexed heterogeneous lists of `PelValue`s, enclosed in brackets `[]`. Example: `[1 "two" #t #nil]`.
- `PelKey`: Represents keyword symbols, prefixed with a colon `:`. Example: `:name`, `:age`. Keywords are often used for named arguments and creating map-like structures.
- `PelPair`: Represents a key-value association, typically formed implicitly by the parser when a `PelKey` is followed by a non-keyword value within a list. Example: `:name "Pel"` within `[:name "Pel" :version 1]`. A standalone key like `:flag` is implicitly paired with `#nil`.
- `PelClosure`: The internal representation of all callable entities (functions and lambdas) in Pel.

## 4.3 Functions and Closures (PelClosure)

Functions are central to Pel.

- **Uniformity:** There are no "special forms" in the traditional Lisp sense. Constructs like

`if`, `for`, `case`, and even `def` are implemented as functions (specifically, `PelClosures`).

- **Definition:** Functions and variables are defined using the same `def` construct. Defining a variable binds a symbol to a value; defining a function binds a symbol to a `PelClosure` (typically created via a `lambda`).

  ```
  (def pi 3.14)          ; Variable definition
  (def greet (lambda [:name] (print "Hello, " name))) ; Function definition
  ```

Thanks to Pel pipes, one can also define functions in "reverse" order—first creating the body and then piping that into a symbol:

```
(lambda [:x :y]
  [(pow x 2) (pow y 2)] ▷ (+) ▷ (sqrt)
 ) ▷ (def my-foo ^)
```

- **PelClosure:** This is the heart of Pel's function mechanism. A `PelClosure` encapsulates:
  - The function's code (or a reference to a built-in implementation).
  - The environment in which it was defined (closing over variables).
  - Specifications for its arguments (`ArgSpec`), including names and optional default values.
  - A flag indicating whether it's `strict` (evaluates arguments before invocation) or `non-strict` (arguments passed unevaluated).
- **Partial Application:** Inspired by Haskell, Pel supports automatic partial application. If a `PelClosure` is called with fewer arguments than required, it does not raise an error. Instead, it returns a *new* `PelClosure` that has captured the provided arguments and is waiting for the remaining ones.

  ```
  (def add (lambda [:x :y] (+ x y))) ; Defines add function
  (def add5 (add 5)) ; Creates a new closure, add5, where x is bound to 5
  (add5 10) ; Invokes the partial closure, returns 15
  ```

  This applies to all functions, including control flow constructs, making them first-class citizens that can be passed around partially applied. For instance, a `for` loop missing its body can be passed as an argument to another function.

- **Strict vs. Non-strict:** Strict functions (the default for lambdas) evaluate their arguments before the function body is executed. Non-strict functions receive their arguments as unevaluated expressions. This is essential for control flow like `if`, which must only evaluate the `then` or `else` branch, not both. Currently, only built-in functions can be non-strict, but user-defined non-strict functions are planned. Non-strict functions in Pel provide an alternative approach to Lisp macros in that they operate directly on the AST and can selectively evaluate their arguments just like the unquote mechanism in macros.

- **Argument Passing:** Closures can be called using:

  - **Positional arguments:** `(area 3 4 5)`
  - **Named arguments:** `(area :x 3 :y 4 :z 5)`. Mixing positional and named arguments in a single call is disallowed. Lambdas can define default values for arguments: `(lambda [:x :y 10] ...)` makes `y` default to `10`.

## 4.4 Piping (▷, ^)

Pel provides a powerful piping mechanism for composing functions linearly, inspired by Elixir.

- **Basic Pipe ▷:** The ▷ operator takes the result of the expression on its left and passes it as the *first* argument to the function call expression on its right.

```
[1 2 3 4] ▷ (len) ▷ (+ 5) ; Equivalent to (+ (len [1 2 3 4]) 5) ⇒ 9
```

- **Caret Injection ^:** The caret symbol `^` acts as a placeholder within the right-hand expression, indicating exactly where the result of the left-hand expression should be injected.

```
"world" ▷ (concat "hello, " ^) ; Equivalent to (concat "hello, " "world")
5 ▷ (* 10 ^) ; Equivalent to (* 10 5)
```

  The caret injection works recursively within nested structures:

```
[1 2 3] ▷ (print :vals ["a list of items:" ^] :sep " ")
; Equivalent to (print :vals ["a list of items:" [1 2 3]] :sep " ")
```

- **Linearity Advantage for LLMs:** This piping mechanism is particularly beneficial for LLMs. They generate code token by token, moving forward. Unlike traditional nested function calls `bar(foo(a))` where the LLM needs to plan the `bar(` call *before* generating `foo(a)`, Pel's pipes allow the LLM to generate `(foo a)`, then decide to pipe it into `bar` *after the fact*: `(foo a) ▷ (bar ^)`. This linear flow aligns better with the sequential nature of LLM generation and reduces the need for complex planning or backtracking.

## 4.5 List Operations and Accessing

Pel provides powerful mechanisms for working with literal lists. While literal lists appear as simple data structures, they actually behave as `PelClosure`s. This design choice enables sophisticated list manipulation operations through a unified function calling syntax.

A literal list is can receive three optional arguments that default to `#nil`: `:at`, `:from`, and `:to`. When a literal list is "called" without arguments, it returns all its elements. However, providing one or more of these arguments enables advanced list slicing and key-value lookup operations:

```
; ① Basic indexing (returns element at index 1)
([5 6 7 8] :at 1) ; => 5
; ② Slicing to a specific index (inclusive)
([5 6 7 8] :to 2) ; => [5 6]
; ③ Slicing from a specific index (inclusive)
([5 6 7 8] :from 2) ; => [6 7 8]
; ④ Combined slicing with both from and to indices
([5 6 7 8] :from 1 :to 3) ; => [5 6 7]
; ⑤ Retrieving multiple elements by index
([5 6 7 8] :at [1 3]) ; => [5 7]
; ⑥ Key-value lookup (requires quoting the key)
```

```
([:a 1 :b 2 :c 3] :at ':a) ; => 1
; ⑦ Multiple key lookup
([:a 1 :b 2 :c 3] :at [':a ':c]) ; => [1 3]
; ⑧ Index-based retrieval of multiple pairs
([:a 1 :b 2 :c 3] :at [1 3]) ; => [:a 1 :c 3]
```

For key lookup operations (as in the sixth example), the key must be quoted to prevent Pel from interpreting it as a separate key-value pair. Without quoting, Pel would interpret `:at :a` as two separate pairs: `PelPair(PelKey(":at"), #nil)` and `PelPair(PelKey(":a"), #nil)`. Literal lists with alternating keys and values are automatically interpreted as containing key-value pairs. Internally, `[:a 1 :b 2 :c 3]` is represented as `[PelPair(PelKey(":a"), PelNum(1)), ...]`. This means standard list operations like indexing still work naturally. Named arguments can be omitted for brevity, with positional arguments being mapped to the list's parameters in order (`:at`, `:from`, `:to`):

```
; Implicit :at parameter (equivalent to :at 1)
([5 6 7 8] 1) ; ⇒ 6
; Implicit :from and :to (equivalent to :from 1 :to 3)
([5 6 7 8] () 1 3) ; ⇒ [6 7 8]
```

Since literal lists behave like closures, they integrate seamlessly with Pel's piping mechanism:

```
(def data [1 2 3 4 5])
(for [0 2 4] i
  i ▷ (data :at ^) ▷ (print))
; Prints elements at indices 0, 2, 4 in data
```

This unified treatment of data structures as closures exemplifies Pel's functional design philosophy and demonstrates how even basic language constructs can offer rich, expressive capabilities through a consistent interface.

## 4.6 Control Flow

Control flow constructs are implemented as non-strict `PelClosures`.

### 4.6.1 `if`

Takes `:cond`, `:then`, and optional `:else` arguments (defaulting to `#nil`). Evaluates `:cond`; if true, evaluates and returns `:then`, otherwise evaluates and returns `:else`.

```
(if data ▷ (len) ▷ (gt 2)
    (print "length of data is greater than 2")
    (print "data is too short"))
```

Note that since even control flow constructs are functions, one can also call them using named arguments and pipe into them:

```
data ▷ (len) ▷ (gt 2) ▷
  (if :cond ^
      :then (print "length of data is greater than 2")
      :else (print "data is too short"))
```

### 4.6.2 `case`

A generalized conditional structure. Takes a value (`:scrut`) and a literal list of condition-consequence pairs (`:body`). It evaluates conditions sequentially. The first condition that evaluates to `#t` causes its corresponding consequence to be evaluated and returned. A final `#t` condition acts as a default `else` clause.

```
(case my-list [
    (len) ▷ (gt 5)
    "length of my-list is greater than 5"
    #t ; Default case
    (print "all conditions failed")
]) ▷ (print)
```

Notice that `case` pipes its `:scrut` into each condition; turning the first condition into `my-list ▷ (len) ▷ (gt 5)`.

### 4.6.3 `for`

Provides iteration. Takes `:coll` (the collection to iterate over), `:iterator` (a symbol to bind each item to), and `:body` (an expression to evaluate for each item). Importantly, it returns a `PelListLiteral` containing the results of each body evaluation. The length of output is the same as `:coll`.

```
(for :coll [1 2 3] :iterator i :body (* i 2)) ; ⇒ [2 4 6]
```

### 4.6.4 `do`

Evaluates a sequence of expressions (provided as arguments or in a `PelListLiteral`) and returns the value of the *last* expression. Useful for side effects.

```
(do
  (print "Starting...")
  (def x 5)
  (+ x 10))
; Prints "Starting...", returns 15
```

### 4.6.5 `do/async`

Similar to `do`, but evaluates the expressions concurrently. Returns the result of the last expression specified in the block, after all have completed.

## 4.7 Natural Language Integration

Recognizing that Pel is often generated or used in conjunction with LLMs, it incorporates direct hooks for LLM evaluation:

- **Natural Language Conditions in `case`:** If a condition expression within a `case` body is a `PelString`, Pel interprets it as a natural language condition. It passes the `:scrut` (the value being tested) and the condition string to an underlying LLM. The LLM's boolean response determines whether the condition passes.

```
(case user-profile [
    "is a premium member" (grant-access user-profile)
    "has incomplete profile" (prompt-completion user-profile)
    #t (show-basic-view)
])
```

Here, the strings are evaluated by an LLM against the `user-profile` data.

- **Other LLM Functions:** Pel can easily incorporate other built-in functions that call LLMs, such as `summarize`, which takes text and returns an LLM-generated summary.

This tight integration allows leveraging the fuzzy understanding capabilities of LLMs directly within the structured logic of Pel.

# 5 The Pel Runtime Environment

Pel is more than just a language specification; it includes a runtime environment designed for interactive development and efficient execution, especially in agentic scenarios.

## 5.1 The REPeL (Read-Eval-Print-Loop)

Pel features an enhanced REPL, affectionately termed "REPeL," which incorporates advanced error handling and debugging features inspired by Common Lisp's condition system and augmented with LLM capabilities. As an interactive development environment, REPeL provides a standard loop for entering Pel code, seeing results, and inspecting the environment. It includes features like command history, syntax highlighting, and auto-completion. But perhaps the most intriguing aspect of REPeL is the way it handles errors. To illustrate this, let us review, as an example, a piece of Pel code that calls two agents, accumulates their responses, and summarizes the result:

```
(FIN-AGENT :query "give me the latest financial report for this quarter"
           :expect "string") ▷ (def financial_report ^)
(SALES-AGENT :query "I need our sales data in detail"
             :expect "string") ▷ (def sales_data ^)
```

```
(add financial_report sales_data) ▷ (summarize) ; erroneous code
```

In this example, two API calls are made to agents, followed by a call to the `add` function. The latter is problematic because the correct way to concatenate two strings in Pel is to use the `concat` function. In such situations, most programming languages (including Python) would throw an exception and exit, discarding the results obtained in the previous lines. Since the API calls could be slow and potentially expensive, we need a way to keep the results obtained from the first two lines while being able to fix the error on the third. Therefore, when any error happens in Pel code (and a `PelException` is raised internally), the REPeL does not simply crash; it preserves the state of the environment *before* the error and presents the user (or an automated system) with restart options:

1. **Rewrite Entire Program:** Discard the current code and enter a completely new program.
2. **Rewrite from Error Forward:** Keep the code up to the error point, discard the rest, and enter new code to replace the faulty part onwards.
3. **Rewrite Current Expression:** Replace only the specific expression that caused the error and retry its evaluation within the original context.
4. **Abort Evaluation:** Stop the current evaluation entirely.
5. **Self-Healing Mode (Helper Agent):** Invoke an LLM-based helper agent to automatically fix the code.

The **LLM-assisted self-healing** is a key innovation of REPeL. When an error occurs, the associated `PelException` captures not only the error message and location but also contextual information, often derived from the docstring of the Pel function where the error originated. A dedicated "Helper Agent" (an LLM prompted with the error, the faulty code snippet, and the function's documentation/context) is invoked. It analyzes the discrepancy between the code and the expected usage (based on the docstring) and proposes a corrected code snippet. In automatic mode, the REPeL can accept this correction and continue execution seamlessly. This provides an "autocorrect" experience, significantly improving robustness, especially when LLMs are generating Pel code.

```
Pel> (def name "Behnam")
     (print ["hello" name] :sep " ")

⇒ "Behnam"
Error at line 2, col 1-32: Mixing named and positional arguments is not allowed.
   1 | (def name "Behnam")
   2 | (print ["hello" name] :sep " ")
       ^^^^^^^^^^^^^^^^^^^^^^^^^^^^^^^^
error context:
    FUNCTION SIGNATURE: (print :vals :sep " " :nl #t)
    TYPES:
        - vals: PelValue - values to print, can be a single value or a literal list
 ↪  [ ... ]
        - sep: PelString (optional) - separator string, default ""
        - nl: PelBool (optional) - whether to end with a newline, default #f
    DESCRIPTION:
        Prints values to stdout. If vals is a bracket-literal, prints each item.
        Optionally separates with the given separator string and adds a newline.
        Returns the input vals unchanged.
    EXAMPLE USAGE:
    <hidden for brevity>

Possible restarts:
1. Rewrite entire program
2. Rewrite from error point forward
3. Rewrite only the current expression
4. Abort evaluation
5. Use self-healing mode
Select option (1-4/5): 5
SELF-HEALING...
Helper agent proposed rewrite:
(print :vals ["hello" name] :sep " ")
Press 'a' to accept, 'e' to edit, 'r' to abort.
Choice (a/e/r)? a
```

```
(print :vals ["hello" name] :sep " ")
```

```
hello Behnam
⇒ ["hello" "Behnam"]
```

## 5.2 Automatic Asynchronicity

Performance is critical for agentic systems where multiple tasks or agent interactions might need to occur concurrently. Pel addresses this through an optional automatic asynchronous execution mode within the REPeL.

- **Dependency Graph Analysis:** Before execution in this mode, the REPeL performs a pre-scan of the entire code. It parses the code into a sequence of top-level Abstract Syntax Trees (ASTs). For each AST, it identifies the symbols it *uses* and the symbols it *defines* (we which specifically look for `def` forms).
- **Parallel Execution:** Based on this analysis, the runtime builds a dependency graph. ASTs that do not depend on the output (defined symbols) of other pending ASTs are considered independent and can be scheduled for concurrent execution using `asyncio` tasks, managed by the `PelTaskManager`. Pel's functional nature and emphasis on immutability simplify this dependency analysis.

This automatic parallelization can significantly speed up Pel programs, especially those involving independent computations or I/O-bound operations like multiple agent calls, without requiring the programmer (or the LLM generating the code) to explicitly manage threads or async primitives everywhere. While this mode offers performance gains, the top-to-bottom execution mode provides more predictable debugging with the restart system; refining error handling in the fully async mode is an area of ongoing work.

# 6 Application: Orchestrating Agentic AI Systems

The primary motivation for Pel was to enable more sophisticated coordination of LLM-based agents. Here we demonstrate how Pel can be leverages to build a hierarchical multi-agent system.

In this example, an organization is modeled as a hierarchy of agents (Figure 3 shows the organizational hierarchy.). For example, there could be a "mini" department of Marketing, a mini department of Finance, etc., each with their own sub-departments. Agent data (roles and backgrounds, available tools for function calling, list of sub-agents) is loaded from a JSON file. In this hierarchy, agents can be designated as **routers** or **terminals**. Terminal agents perform specific tasks (potentially using traditional function calling or direct LLM responses). Router agents, crucially, coordinate their sub-agents.

**Pel as the Orchestration Language:** Instead of relying on complex internal logic or limited function calls, router agents achieve coordination by generating and executing Pel code. A router agent, when tasked by its supervisor, might write Pel code that:

- Calls one or more of its sub-agents sequentially or in parallel (`do/async`).
- Uses the output of one sub-agent as input (`:context`) for another, facilitated by Pel's variables (`def`) and piping (`▷`).
- Implements conditional logic (`if`, `case`) based on sub-agent responses to decide the next steps.
- Initiates collaborative sessions among sub-agents using the built-in `meeting` function, which takes a list of agent names, a topic, and simulates discussion rounds. The transcript can then be processed (e.g., summarized using `summarize`).

For instance, when the user asks the `MAIN` agent to come up with a comprehensive plan for social media advertising, the agent writes the following Pel code which queries the relevant agents (`FINANCE` and `MARKETING`) for more information and aggregates the results in a list. That list, once evaluated, is given back to `MAIN`, which it then uses to provide an answer to the user:

```
; MAIN's Pel code

(MAIN/FINANCE :query "what's the budget allocation
                      for social media advertising?"
              :expect "num")
  ▷ (def social_media_budget ^)
```

```
(MAIN/MARKETING :query "come up with an effective advertising campaign
                        on social media given the given budget"
                :context social_media_budget
                :expect "string")
  ▷ (def social_media_strategy ^)

[:social_media_budget social_media_budget
 :social_media_strategy social_media_strategy]
```

Notice that the call to `MARKETING` itself might trigger that agent to write a piece of Pel code to orchestrate its subagents:

```
; MARKETING's Pel code

(meeting :group ["MAIN/MARKETING/SOCIAL_MEDIA"
                 "MAIN/MARKETING/CONTENT_MARKETING"]
         :rounds 3
         :topic "come up with a great advertising campaign given the budget"
         :context social_media_budget)
  ▷ (summarize)
  ▷ (def plan_summary ^)

[:plan_summary plan_summary]
```

This Pel-based approach provides far greater expressiveness and flexibility than function calling. Complex, multi-step, conditional workflows involving multiple agents can be explicitly defined and executed. Compared to generating Python, Pel offers inherent safety benefits through its controlled grammar and avoids the need for complex sandboxing. Moreover, if a router agent generates syntactically incorrect or semantically flawed Pel code (e.g., calls a non-existent sub-agent, uses a function incorrectly), the REPeL's restart mechanism kicks in. In an automated setup (with self-healing), the helper agent can often fix the Pel code, allowing the agentic system to recover gracefully from errors that would halt systems relying on less robust execution models.

Pel thus serves as the connective tissue, the "language of thought and action," for these hierarchical agentic systems, enabling complex emergent behaviors through structured, safe, and expressive code.

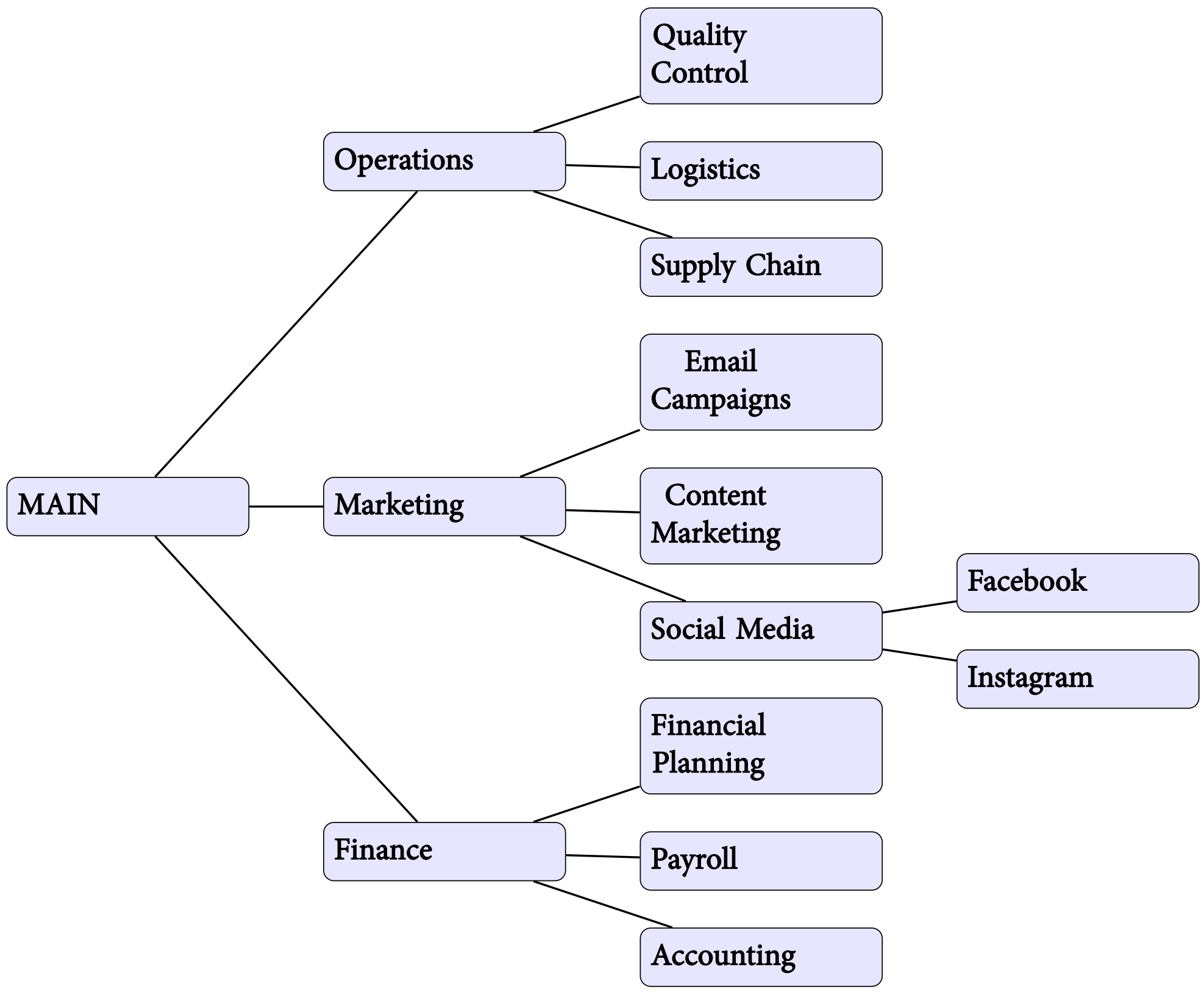

Figure 3: Organizational structure diagram

# 7 Discussion and Future Work

This paper introduces Pel, a specialized language designed specifically for Large Language Model (LLM) orchestration. Pel provides a dedicated, secure, and expressive environment for LLM code generation that successfully navigates between the limitations of basic function calling and the security risks of unrestricted general-purpose language generation.

At its core, Pel leverages a simple, Lisp-inspired grammar that enables fine-grained control over LLM-generated code capabilities at the syntax level. This design choice allows for grammar modification and constrained generation, substantially reducing security risks without requiring complex sandboxing mechanisms. Despite this syntactic simplicity, Pel achieves enhanced expressiveness by incorporating control flow constructs (if, case, for) as first-class functions and implementing a powerful piping mechanism inspired by Elixir and Gleam. These features enable complex, linearly-composed workflows that are particularly well-suited for LLM generation.

A key innovation in Pel is its seamless integration with LLMs, natively supporting natural language conditions within its control flow and delegating their evaluation to LLMs when necessary. This integration extends to Pel's advanced development environment (REPeL), which features an interactive REPL with Common Lisp-style restarts and LLM-powered "Helper Agents" for automated error diagnosis and correction. Furthermore, Pel's runtime can automatically detect and parallelize independent code blocks by analyzing dependencies in the Abstract Syntax Tree (AST), enhancing performance for agentic systems.

Pel emerges from both practical necessity and appreciation for powerful programming language paradigms, offering a novel solution to the critical challenge of LLM orchestration and agency. By rejecting the false dichotomy between overly simplistic function calling and insecure general-purpose code generation, Pel establishes a middle ground: a language specifically designed for LLMs to use safely and effectively. As LLMs become increasingly integrated into complex workflows and autonomous systems, languages like Pel, which bridge the gap between natural language understanding and structured, safe execution, will be essential. Pel offers a promising foundation for building more capable, reliable, and controllable AI agents, representing a significant step toward specialized languages for interacting with and controlling LLMs. Its core contributions lie in its unique combination of syntactic simplicity, grammar-level safety, expressive pipeline-style programming, and advanced error recovery in its REPeL through self-healing agents and a robust restart system.

## 7.1 Limitations and Future Directions

Currently, only built-in functions can be non-strict, but enhancing expressiveness by allowing users to define their own non-strict functions remains a significant area for improvement, albeit one that requires careful semantic design. Additionally, work is ongoing to improve the predictability and usability of the restart mechanism within the fully automatic asynchronous error handling mode. While Pel has demonstrated impressive few-shot learnability, another promising direction involves fine-tuning LLMs, even small models, specifically for Pel code generation, which could substantially improve code quality and reduce errors. Future work could concentrate on addressing these areas for improvement.